\documentstyle[12pt]{article}

\advance\hoffset by -1.1truecm
\advance\voffset by -1.0truecm

\def\be{\begin{equation}}
\def\ee{\end{equation}}
\def\ba{\begin{eqnarray}}
\def\ea{\end{eqnarray}}
\def\v#1{\vert #1 \rangle}
\def\scp#1#2{\langle#1\vert#2\rangle}

\newcommand{\x}{{\bf x}}
\newcommand{\p}{{\bf p}}
\newcommand{\A}{{\cal A}}

\newcommand{\sn}{\smallskip\newline}
\newcommand{\mn}{\medskip\newline}

\newcommand{\mbo}{{\mbox{ }}}

\def\sec#1{\vskip0.5cm { \large \bf {\flushleft #1} \rm }\sn    }

\begin{document}

\title{On Noncommutative Geometric Regularisation}

\author{Achim Kempf\thanks{ Research Fellow of Corpus Christi College in the
University of Cambridge}\\ 
Department of Applied Mathematics \& Theoretical Physics\\
University of Cambridge, Cambridge CB3 9EW, U.K.\\
 {\small Email: a.kempf@amtp.cam.ac.uk}}

\date{}
\maketitle

\vskip-7.2cm
{\tt $\mbo$ \hskip11.1cm DAMTP/96-23}\newline
{\tt $\mbo$ \hskip11.72cm  hep-th/9602119}\rm
\vskip6.9cm

\begin{abstract}
Studies in string theory and in quantum gravity suggest the existence 
of a finite lower bound to the possible resolution of lengths  which, 
quantum  theoretically,  takes the form  of a minimal uncertainty  in 
positions  $\Delta x_0$.  A finite  minimal  uncertainty  in  momenta 
$\Delta p_0$  has  been  motivated from the absence of plane waves on 
generic curved spaces.   Both  effects  can  be  described  as  small 
noncommutative geometric  features of space-time.  In a path integral 
approach  to  the  formulation  of  field  theories on noncommutative 
geometries,  we  can  now  generally  prove IR regularisation for the 
case  of  noncommutative geometries which imply minimal uncertainties
$\Delta p_0$ in momenta. 
\end{abstract}

\sec{1. Introduction}
As has been known for long, 
for the resolution of small distances  
high energetic test particles are needed 
which through their gravitational effect 
will disturb eventually significantly
the spacetime structure which was tried to be resolved.
The problem has been approached from several directions and studies
in string theory and quantum gravity suggest that, quantum theoretically,
a lower bound to the resolution of distances 
could take the form of a finite minimal position
uncertainty $\Delta x_0$
of the order of the Planck length of $\approx 10^{-35}m$, 
see e.g. \cite{townsend}-\cite{garay}.
On the other hand, on large scales, there is no notion of plane waves
or momentum eigenvectors on generic curved spaces. It has therefore 
been suggested that quantum theoretically
there could then exist lower bounds
$\Delta p_0$ to the possible determination of momentum \cite{ft,ct}.
\sn
Independent of the suggested mechanisms for the 
origins of minimal uncertainties (or whether they are intended 
as a formal regularisation only)
both types of effects, 
i.e. a $\Delta x_0$ or a $\Delta p_0$ can be 
described as small noncommutative geometric corrections to 
space-time and/or energy-momentum space 
\cite{ixtapa}-\cite{ak-hh-1}. 
\sn
Intuitively, the presence of finite minimal uncertainties
$\Delta x_0,\Delta p_0$ should
have UV and IR regularising effect in field theory. 
The example of euclidean $\phi^4$-theory on a restricted 
class of such noncommutative geometries has been studied in detail
and both UV and IR regularisation have been shown for this case
\cite{ft}-\cite{ak-jmp-qft}.
\mn
Our aim here is to 
prove the IR regularity of euclidean propagators $1/(\p^2 +m^2c^2)$
for all noncommutative geometries with a minimal uncertainty in momentum 
$\Delta p_0$, both for $m>0$ and for $m=0$.

\sec{2. Noncommutative geometries with minimal uncertainties}
We consider the possibility of small `noncommutative
geometric' corrections to the canonical commutation relations
in the associative 
Heisenberg algebra ${\cal{A}}$ generated by the $\x_i,\p_j$, see
\cite{ixtapa}-\cite{ak-jmp-qft}:
\be
[\x_i,\p_j] = i\hbar ( \delta_{ij} + \alpha_{ijkl} \x_k\x_l +
\beta_{ijkl} \p_k\p_l + ... )
\label{ba}
\ee
and also
\be 
[\x_i,\x_j] \ne 0, \mbox{\quad \quad \quad} [\p_i,\p_j] \ne 0 
\label{ab}
\ee
with the involution $\x_i^* = \x_i, \p_i^* = \p_i$.
\sn
A priori we formulate field theories on
generic noncommutative background `geometries' $\A$ which may or
may not have certain symmetries, similar to the 
case of curved background
geometries. While 
nontrivial examples of non Lorentz 
symmetric noncommutative background geometries 
have been studied in \cite{ixtapa}-\cite{ak-jmp-qft},
Lorentz symmetric examples 
of suitable noncommutative background geometries 
were found in \cite{ak-gm-rm-prd}.
\sn
The correction terms necessarily imply new physical features,
since unitary transformations
are generally commutation relations preserving. Here,
for appropriate small $\alpha,\beta$
one obtains ordinary quantum mechanical behaviour at medium scales
while the presence of small $\alpha$ and $\beta$ 
imply modified IR and UV behaviour respectively:
\sn
The uncertainty relations, holding in all $*$-representations 
of the commutation relations on some 
dense domain $D \subset H$ in a Hilbert space $H$,
are of the form $ \Delta A \Delta B \ge 1/2 \vert\langle[A,B]\rangle\vert $
so that e.g. $[\x_i,\x_j]\ne 0$, yields $\Delta x_i \Delta x_j \ge 0$. 
The noncommutativity implies that the $\x_i$ (as well as the 
$\p_i$) can no longer be simultaneously diagonalised.
Because of Eqs.\ref{ba} and
the corresponding uncertainty relations there can appear the even more
drastic effect that the $\x_i$ (as well as the $\p_j$) may also not be 
diagonalisable separately.
\sn
Already in one dimension the uncertainty relation
(assuming small positive $\alpha,\beta$ with 
$\alpha \beta < 1/\hbar^2$ and neglecting higher order corrections):
\begin{equation}
\Delta x \Delta p \ge \frac{\hbar}{2} \left( 1 + \alpha (\Delta x)^2 
+ \alpha \langle \x\rangle ^2
+ \beta (\Delta p)^2 + \beta \langle \p\rangle ^2 \right)
\label{1dimucr}
\end{equation}
implies nonzero minimal uncertainties in $\x$- as well as in $\p$- 
measurements:
$ \Delta x_{0} = (1/\beta\hbar^2 - \alpha)^{-1/2}, \quad
\Delta p_{0} = (1/\alpha\hbar^2 - \beta)^{-1/2}$.
For $\alpha=0$ and a small $\beta$ we cover the example of an 
ultraviolet modified uncertainty
relation that has been discussed in string theory and 
quantum gravity, for a review see \cite{garay}.
For all physical domains $D$, i.e. for all
$*$-representations of the commutation relations, there 
are now no physical states in the minimal uncertainty `gap'
\be
\exists\!\!\!/ \quad \vert \psi \rangle \in D: \quad
0 \le (\Delta x)_{\vert \psi \rangle} < \Delta x_0
\ee
\be
\exists\!\!\!/ \quad \vert \psi \rangle \in D: \quad
0 \le (\Delta p)_{\vert \psi \rangle} < \Delta p_0
\ee
where $\v{\psi}$ generally stands for a normalised element of $D$.  
Thus, unlike on ordinary geometry, there now 
do not exist sequences $\{\vert \psi_n \rangle\}$ of 
physical states which would approximate point localisations in 
position or momentum space, i.e. for which
the uncertainty would decrease to zero, e.g.
$ \exists\!\!\!/ \quad \v{\psi_n} \in D: \quad 
\lim_{n\rightarrow \infty} (\Delta x)_{\vert \psi_n \rangle} = 0$.
\sn
Technically, the new infrared and ultraviolet behaviour has important
consequences for the representation theory, in that e.g. a 
finite minimal uncertainty $\Delta x_0$ in positions implies 
that the commutation
relations do no longer find a spectral representation of $\x$, so
that one has to resort to other Hilbert space representations.
\sn
The interplay between the 
functional analysis of the position and the momentum operators 
has first been studied in \cite{ixtapa,ak-jmp-ucr}. In fact we are giving up 
(essential) self-adjointness of the $\x$ and $\p$ operators, to retain only
their symmetry. While giving up essential self-adjointness
is necessary for the description of the
new short distance behaviour, the symmetry is sufficient to guarantee that all physical expectation 
values are real and also that uncertainties can be calculated
applying the usual definition of the standard deviation,
e.g. $\Delta x = \langle\psi\vert(\x
-\langle\psi\vert\x\vert\psi\rangle)^2\vert\psi\rangle^{1/2}$.
Nevertheless, this is a nontrivial step which
goes beyond the conventional quantum mechanical treatment, and it 
also goes beyond Connes' `dictionary' \cite{connes}
of how to treat `real variables' on noncommutative geometries. 
\sn
The key observation is that although
self-adjoint extensions and (discrete) diagonalisations of
$\x$ or $\p$ exist in $H$, 
under the circumstances described, these 
diagonalisations are not in any common domain, i.e. not in any physical
domain, of $\x$ \it and \rm $\p$ \cite{ixtapa,ak-jmp-ucr}.
Instead there is now the finite
uncertainty `gap' separating the physical states from formal $\x$
or $\p$ eigenstates. For the details and proofs 
see \cite{ixtapa,ak-jmp-ucr,ak-jmp-qft,ak-gm-rm-prd,ak-hh-1}. 
\sn
Concerning the infrared we remark that due to the correction terms 
the momenta $\p_i$ no longer
generate translations on flat space. Under certain conditions,
the $\p_i$ do generate translations of normal coordinate frames
on curved spaces, see \cite{ft,ct}, in which case the relation between the
absence of plane waves (i.e. of $\p$-eigenstates) and the presence of a
minimal uncertainty in momentum can be explicitly investigated.
\sn
The physical states of then maximal 
localisation have in the meanwhile been extensively 
studied, first in the special case $\alpha = 0,\beta>0$, 
see \cite{ak-gm-rm-prd}
and recently also in the general (though one-dimensional) 
case $\alpha,\beta>0$, see \cite{ak-hh-1}.
Explicitly, the physical states 
$\v{\phi^{mlx}_{\xi}}$, $\v{\phi^{mlp}_{\pi}}$
which realise the maximal localisation 
in positions or momenta obey
\be
\Delta x_{\v{\phi^{mlx}_{\xi}}} = \Delta x_0
\mbox{ , \qquad  }
\langle \phi^{mlx}_{\xi} \vert \x \v{\phi^{mlx}_{\xi}} = \xi
\mbox{ , \qquad \quad }
\langle \phi^{mlx}_{\xi} \vert \p \v{\phi^{mlx}_{\xi} } = 0
\ee
and similarly for $\v{\phi^{mlp}_{\pi}}$.
E.g. the projection $\scp{\phi^{mlx}_{\xi} }{\psi}$ is then the
probability amplitude for finding the particle maximally localised
around $\xi$. For $\alpha, \beta  \rightarrow 0$ one recovers the 
position and the momentum eigenvectors. n-dimensional
studies are in progress, we here only state one key result, the 
mutual projection of maximal localisation states:
\be
\langle \psi^{mlx}_{{\xi}^{\prime}}
 \vert \psi^{mlx}_{\xi} \rangle 
= \frac{1}{\pi} \left( 
\frac{{\xi}-{\xi}^{\prime}}{2\hbar\sqrt{\beta}}  
 - \left(\frac{{\xi}-{\xi}^{\prime}}{2\hbar 
 \sqrt{\beta}}\right)^3\right)^{-1}
 \sin\left(\frac{{\xi}-{\xi}^{\prime}}{2\hbar\sqrt{\beta}}\pi\right)
\label{nd}
\ee
It is the generalisation of the Dirac
$\delta$-function which on ordinary geometry 
would be obtained 
{}from projecting maximal localisation states, i.e.
then from projecting position eigenstates 
onto another: $\langle x\vert x'\rangle
= \delta(x-x')$.
The nonmultiplicativity of $\delta$- distributions is related to
the appearance of ultraviolet divergencies, whereas the behaviour 
of Eq.\ref{nd} (note that the singularities of its first factor 
are cancelled by zeros of the sinus)
suggests UV regularity in field theory.

\sec{3. Path Integration} 
The ansatz \cite{ak-jmp-ucr}-\cite{ak-jmp-qft}
for the formulation of field theories on
such noncommutative geometries is explained most easily in 
the simple example of charged euclidean $\phi^4$-theory: The
partition function
\be
Z[J] := N \int D\phi\mbo e^{\int d^4x\mbo
\phi^* (\partial_i\partial_i - \mu^2)\phi 
 - \frac{\lambda}{4!}(\phi \phi)^*\phi \phi + \phi^*J+J^*\phi}
\label{pfx}
\ee
we write in the form 
\be
Z[J] = N \int_D D\phi\mbo e^{- \bf tr\rm\left( \frac{l^2}{\hbar^2}
(\p^2+m^2c^2).\vert\phi\rangle\langle\phi\vert \mbo - 
\mbo \frac{\lambda l^4}{4!}
\vert\phi *\phi\rangle\langle\phi *\phi\vert \mbo + 
\mbo \vert\phi\rangle\langle J\vert
\mbo + \mbo \vert J\rangle\langle\phi\vert\right)}
\label{pf}
\ee
where, to make the units transparent, 
we introduced an arbitrary positive length to render
the fields unitless ($l$ could trivially be reabsorbed in the fields).
\sn
Eq.\ref{pfx} is recovered from Eq.\ref{pf} by assuming the ordinary
relations $[\x_i,\p_j]= i \hbar \delta_{ij}$ in $\A$ and by choosing 
the spectral representation of the $\x_i$. We then have as usual 
$\phi(x) := \langle x\vert \phi\rangle$ with the scalar product
$\langle\phi\vert\psi\rangle=\int d^4x \phi^*(x) \psi(x)$,
i.e. the trace $
\bf tr\rm(q) = \int d^4x \mbo \langle x\vert q\vert x\rangle
$, and the operators acting as
$
\x_i.\phi(x) = x_i\phi(x), \mbo \p_i.\phi(x) = -i\hbar\partial_{x_i}\phi(x)
$.
\sn
The pointwise multiplication $*$
\be
(\phi_1 * \phi_2)(x) = \phi_1(x) \phi_2(x), \mbox{ \quad i.e. \quad }
\langle x\vert \phi_1*\phi_2\rangle = \langle x\vert \phi_1\rangle\langle 
x\vert \phi_2\rangle
\ee
which expresses point interaction, 
is (and can also on noncommutative geometries be kept) 
commutative for bosons.
Since fields are in a representation of $\A$, similar to quantum mechanical 
states, we formally extended Dirac's 
bra-ket notation for states to fields. In Eq.\ref{pf}
this yields a convenient notation for the functional analytic structure 
of the action functional, but of course, the quantum mechanical
interpretation does not simply extend, see e.g. \cite{fdl}.
The space $D$ of fields that is formally to be summed over can be taken
to be the dense domain $S_{\infty}$ in the Hilbert space $H$ 
of square integrable fields.                 
\sn
Generally, the unitary transformations that map from one
Hilbert basis to another have trivial determinant, so that no anomalies are
introduced into the field theory and changes of basis can be
performed arbitrarily,
in the action functional, in the Feynman rules or in the end results
of the calculation of $n$- point functions.
\sn
Let us now assume that the commutation relations, 
i.e. ${\cal A}$, are represented on a dense domain $D$ spanned
by a Hilbert basis of vectors $\{\v{n}\}$, where $n$ may
be discrete, as e.g. in the case of a Bargmann Fock representation or
it may be continuous, as in the case e.g. of position or
momentum representations, or it may generally be a mixture of both.
Keeping this in mind we use the notation for $n$ discrete.
The identity operator on $H$ can then be written 
$1 = \sum_n \vert n \rangle \langle n \vert$ and fields and operators
are expanded as 
\be
\phi_n = \langle n\vert \phi \rangle \qquad 
\mbox{and e.g. }\qquad 
(\p^2 +m^2c^2)_{nn^{\prime}} = \langle n \vert \p^2 + m^2c^2 \vert
n^{\prime}\rangle
\label{pb}
\ee
The pointwise multiplication, which we will need for the local  
interaction then reads
\be
* = \sum_{n_i} L_{n_1,n_2,n_3} \vert n_1\rangle \otimes
\langle n_2\vert \otimes \langle n_3 \vert
\ee
In this Hilbert basis the partition function reads, summing 
over repeated indices:
\be
Z[J] = N \int_D D \phi \mbo e^{ -\frac{l^2}{\hbar^2}\mbo
\phi_{n_1}^* (\p^2 +m^2c^2)_{n_1n_2} \phi_{n_2}
-\frac{\lambda l^4}{4!} L^*_{n_1n_2n_3}
L_{n_1n_4n_5} \phi^*_{n_2} \phi^*_{n_3} \phi_{n_4}\phi_{n_5}
+ \phi^*_n J_n + J^*_n \phi_n }
\label{d1}
\ee
Pulling the interaction term in front of the path integral,
completing the squares, and carrying out the gau{\ss}ian integrals
yields 
\be
Z[J] = N' e^{-\frac{\lambda l^4}{4!} L^*_{n_1n_2n_3} L_{n_1n_4n_5}
\frac{\partial}{\partial J_{n_2}}
\frac{\partial}{\partial J_{n_3}}
\frac{\partial}{\partial J^*_{n_4}}
\frac{\partial}{\partial J^*_{n_5}}}
\mbo
e^{-\frac{l^2}{\hbar^2} J^*_n (\p^2 +m^2c^2)^{-1}_{nn^{\prime}} 
J_{n^{\prime}}}
\ee
The inversion of $(\p^2 +m^2c^2)$ is nontrivial and involves
a self-adjoint extension in which it can be 
diagonalised and inverted. This will be investigated below.
We obtain the Feynman rules:
\be
\Delta_{n_1n_2} = \left(\frac{-\hbar^2}{l^2(\p^2
+m^2c^2)}\right)_{n_1n_2}
\qquad
\mbox{and } \qquad
\Gamma_{n_1n_2n_3n_4} = -\frac{\lambda l^4}{4!}
L^*_{n^{\prime}n_1n_2} L_{n^{\prime}n_3n_4} 
\label{d2}
\ee
Note that since each vertex attaches to four propagators,
$l$ drops out of the Feynman rules, as it should be.\newline
Let us recall that the usual formulation
of partition functions, such as e.g. Eq.\ref{pfx}, implies that 
$\p^2$ can be represented as the Laplacian on a spectral representation 
of the $\x_i$, so that $\p_i$ is represented
as $-i\hbar\partial_i$, i.e. it is implied 
that $ [\x_i,\p_j] = i\hbar \delta_{ij}$.
It is crucial that in our formulation of 
partition functions in abstract form, such as in
Eq.\ref{pf}, the commutation
relations of the underlying algebra $\A$ are not implicitly fixed
and can be generalised, e.g. to the form of 
Eqs.\ref{ba},\ref{ab}.

Representing $\A$ on some dense $D$ in a Hilbert space $H$
with an e.g. discrete Hilbert basis $\{\v{n}\}$ (the 
Hilbert space is separable), one straightforwardly obtains the
Feynman rules through 
Eqs.\ref{pf},\ref{pb}-\ref{d2}. Thus, in particular, 
the formalism allows to explicitly check noncommutative geometries on  
UV and IR regularisation. 

\sec{4. IR Regularisation}
On ordinary geometry a finite mass term in the propagator 
$(\p^2+m^2c^2)^{-1}$ ensures that, as an operator, it is bounded
(we need not specify a Hilbert basis, as we did e.g. in
Eq.\ref{pb}). 
However, for $m=0$ the operator $1/\p^2$ is unbounded, 
causing infrared divergencies. Indeed,  
on geometries that imply a minimal uncertainty $\Delta p_0$ 
even the massless propagator $1/\p^2$ 
is as well behaved as if it contained a mass term.
\sn
To be precise, we intend to show that for all noncommutative 
geometric algebras $\A$ of the type of Eqs.\ref{ba},\ref{ab}
which imply a minimal uncertainty $\Delta p_0$ there holds for
$m>0$ as well as for $m=0$:
\mn \it
\bf A \rm \quad \it The operator $(\p^2 +m^2c^2):= (\sum_i \p_i\p_i + m^2c^2)$
has exactly one 
self-adjoint extension 
$(\p^2 +m^2c^2)_F$ which is contained in its form domain. \rm
\sn
\bf B \rm \quad \it The operator $(\p^2 +m^2c^2)_F$ has a unique inverse
(the free propagator) which is self-adjoint and defined on the entire
Hilbert space H.    \rm
\sn
\bf C \rm \quad \it The propagator $(\p^2 +m^2c^2)^{-1}_F$ is 
infrared 
regular, i.e. it is a bounded operator (implying also that
its matrix elements are bounded) with bound
$ \vert\vert (\p^2+m^2c^2)_F^{-1}\vert\vert \le
(n(\Delta p_0)^{2} + m^2c^2)^{-1}$. \rm
\sn
\bf D \rm \quad \it Also propagators that are the inverse to 
arbitrary
other self-adjoint extensions of $(\p^2 +m^2c^2)$ (for finite deficiency
indices) are IR-regular,  i.e. they are 
bounded self-adjoint operators on $H$. \rm
\mn
To see this, let $\A$ be represented on a dense domain $D\subset H$ in
a Hilbert space $H$.
By assumption the momenta $\p_i$ exhibit 
a minimal uncertainty $\Delta p_0>0$, i.e. for all states, i.e for all
\it normalised \rm vectors $\v{\phi} \in D$ holds
$\Delta p_i{}_{\v{\phi}}\ge \Delta p_0$, so that
\be
\langle \phi \vert \p_i^2 \vert \phi \rangle =\langle\phi\vert\p_i\vert
\phi\rangle^2+({\Delta p_i}_{\v{\phi}})^2 \ge (\Delta p_0)^2
\ee
and by linearity, for vectors of arbitrary norm:
\be
\langle\phi \vert\p^2\vert\phi\rangle \ge n \vert\vert \phi\vert\vert^2 
(\Delta p_0)^2
\label{sb}
\ee
Thus, the operator $(\p^2+m^2c^2)$ is
a densely defined symmetric \it positive definite \rm operator
(now even for $m=0$), and therefore has,
by a theorem of Friedrich, see e.g.\cite{gbook}-\cite{reedsimon}, a unique 
self-adjoint extension within its form domain. It has the same lower
bound as the original operator.
Explicitly, the Friedrich extension $(\p^2+m^2c^2)_F$ 
of $(\p^2+m^2c^2)$ has the domain $D_F = D_{(\p^2+m^2c^2)^*} \cap H^\prime$,
which is the intersection of the domain 
$D_{(\p^2+m^2c^2)^*}$ of the adjoint $(\p^2+m^2c^2)^*$ with the Hilbert space
$H^\prime$ obtained by completion of $D$ with respect to the norm
$\vert\vert \phi \vert\vert^\prime := 
\langle \phi\vert \p^2 +m^2c^2 
\vert\phi\rangle^{1/2}$ induced by the quadratic form
which is defined through the positive definite 
operator $(\p^2+m^2c^2)$.
The range of $(\p^2+m^2c^2)_F$ is
$R((\p^2+m^2c^2)_F) = H$, the inverse
$(\p^{2}+m^2c^2)_F^{-1}$ exists, 
has the domain $D_{(\p^{2}+m^2c^2)_F^{-1}}= R((\p^2+m^2c^2)_F)=H$,
and is a self-adjoint bounded operator: 
\be
\vert\vert (\p^2+m^2c^2)_F^{-1}\vert\vert \le
\frac{1}{n(\Delta p_0)^{2} + m^2c^2}
\ee
For a constructive proof of the properties of the Friedrich
extension see e.g.\cite{maurin}.
To see the invertibility note that,
since $(\p^2+m^2c^2)_F$ has the same bound as $(\p^2+m^2c^2)$, i.e.
$\forall \v{\phi} \in D_F: \langle
\phi\vert(\p^2+m^2c^2)_F\vert\phi\rangle \ge 
m^2c^2 + n \vert\vert \phi\vert\vert^2 
(\Delta p_0)^2$, its kernel is empty:
$(\p^2+m^2c^2)_F\v{\phi}=0 \Rightarrow 0= 
\langle \phi\vert (\p^2+m^2c^2)_F\vert\phi\rangle
\ge m^2c^2 + n \vert\vert \phi\vert\vert^2 (\Delta p_0)^2$. 
Due to the Cauchi Schwarz inequality also the matrix elements
of $(\p^2+m^2c^2)_F^{-1}$ are bounded:
\begin{eqnarray}
\forall \v{\phi},\v{\psi}\in H: \quad
\vert\langle\phi\vert (\p^2+m^2c^2)_F^{-1}\vert\psi\rangle\vert & \le &
\vert\vert \phi\vert\vert \mbo \vert\vert\psi\vert\vert\mbo
\vert\vert (\p^2+m^2c^2)_F^{-1}\vert\vert \nonumber \\
  & \le &
\vert\vert \phi\vert\vert \mbo \vert\vert\psi\vert\vert\mbo
(n (\Delta p_0)^{2} + m^2c^2)^{-1}
\end{eqnarray}
So far we have shown {\bf A}-{\bf C}, i.e. 
that there exists a canonical inverse $(\p^{2}+m^2c^2)_F^{-1}$ 
and that, as a
propagator, it does not lead to infrared problems, since it is bounded,
also in the case $m=0$.
\sn
To see {\bf D} we consider the bi-adjoint $(\p^2+m^2c^2)^{**}$, which
is symmetric and closed, as is every bi-adjoint of a densely defined
symmetric operator. Due to the existence of one self-adjoint
extension, $(\p^2+m^2c^2)_F$, the deficiency indices $(r,r)$ are equal.
\sn
We recall that on ordinary geometry the deficiency indices are $(0,0)$,
implying that $(\p^2+m^2c^2)_F$ is the only self-adjoint extension.
The deficiency indices can now be nonzero, examples of which 
are known, see \cite{ixtapa,ak-jmp-ucr,ak-gm-rm-prd,ak-hh-1}.
There then exists 
a whole family of further self-adjoint extensions $(\p^2+m^2c^2)_f$ 
(e.g. labeled by $f$), and a corresponding 
 family of propagators $(\p^2+m^2c^2)_f^{-1}$ 
(for invertible $(\p^2+m^2c^2)_f$) which, in explicit representations,
differ by their boundary conditions. 
\sn
A priori we do not want to exclude
these nonstandard propagators (although we exclude as unphysical
the case of infinite
deficiency indices in which case the propagator would require an infinite
set of boundary conditions).
\sn
Indeed, also the nonstandard propagators are IR regular.
To see this, we note first that also
$(\p^2+m^2c^2)^{**}$ is  semibound
{}from below by ($n(\Delta p_0)^2 + m^2c^2$) since $(\p^2 +m^2c^2)_F$ which
is an extension of $(\p^2+m^2c^2)^{**}$ has this property.
As seen by the v. Neumann method, the unitary extension of the
isometric Cayley transform only involves a finite dimensional
mapping of the deficiency spaces and thus 
all self-adjoint extensions of a closed symmetric operator
have the same essential spectrum, see e.g. Thm.8.18 in \cite{gbook}.
Indeed, since $(\p^2+m^2c^2)^{**}$ is closed, symmetric and
bounded from below the now interesting part of the spectrum 
$\sigma((\p^2+m^2c^2)_f)\cap(-\infty,
n(\Delta p_0)^2 +m^2c^2)$ of its self-adjoint extensions
consists of isolated eigenvalues only, of
total multiplicity $\le r$, 
see e.g. Cor.2 of Thm8.18 in \cite{gbook}. Thus, for all invertible
self-adjoint extensions there exist $\epsilon >0$ so that
the spectrum is empty in the finite
intervall $[-\epsilon,\epsilon]$ i.e. in the neighbourhood of zero. 
We can therefore conclude
the boundedness of the spectra of the inverses to 
arbitrary invertible self-adjoint extensions
of $(\p^2+m^2c^2)$.
To be precise, for all invertible self-adjoint extensions
zero is a regular point $0\in \rho((\p^2+m^2c^2)_f)$
since it is not in the spectrum. For self-adjoint
operators $A$ there generally holds, see e.g. 
\cite{gbook} (Thm5.24), \cite{smirnov} (Thms129.1,2) or
\cite{ag}-\cite{reedsimon}:
\begin{eqnarray}
z\in\rho(A)  & \Leftrightarrow & \exists c>0,\forall v\in D(A):
 \vert\vert (z-A).v \vert\vert \ge c \vert\vert v\vert\vert \mbo 
(\mbox{and} \Rightarrow
\vert\vert1/(z-A)\vert\vert \le c^{-1}) \nonumber \\
  & \Leftrightarrow & R(z-A) = H 
\end{eqnarray}
Here, we therefore have:
\be
\exists \epsilon > 0, \forall v\in D_f: \quad
\vert\vert (\p^2+m^2c^2)_f.v\vert\vert \ge \epsilon \vert\vert v \vert\vert
\mbo \mbox{ and } \mbo R((\p^2+m^2c^2)_f)=H
\ee
Thus, the corresponding propagators are bounded
$\vert\vert (\p^2+m^2c^2)^{-1}_f\vert\vert \le 1/\epsilon$ and
are defined on the entire
Hilbert space $D_{(\p^2+m^2c^2)_f^{-1}} = R((\p^2+m^2c^2)_f)=H$.
Also, the propagators are 
self-adjoint, as the inverses of self-adjoint operators
generally are.
\sec{5. Outlook}
Concerning the ultraviolet,  
the same arguments prove of course that e.g. a background
Coulomb potential $A_\mu(\x) := (q/\sqrt{(\sum_i\x_i^2)_F},0,0,0)$ 
(the square root is well defined since $(\sum_i\x_i^2)_F$ is
positive definite) is bounded
in the presence of a minimal uncertainty $\Delta x_0$ in positions.
Given a representation of the algebra $\A$, the propagator
$\Delta = ((\p_i+e A_i(\x))^2+m^2c^2)_F)^{-1}$ can 
be calculated straightforwardly.
Investigations into the `local' gauge principle on
geometries with minimal uncertainties should eventually
allow to study also dynamical
gauge fields and to check for UV regularisation. 
While in the simpler $\phi^4$-theory detailed studies
have been carried out for certain classes of geometries
with minimal
uncertainties $\Delta x_0$ and $\Delta p_0$, \cite{ft}-\cite{ak-jmp-qft}, 
let us here only remark that for UV-regularisation 
the structure of the pointwise
multiplication $*$ that describes local
interaction is crucial. Due to the absence of a position representation,
 $*$ is nonunique in the case of $\Delta x_0 > 0$.
Crucially, an interaction is now
observationally local if any formal nonlocality of $*$ is not larger than
the scale of the nonlocality $\Delta x_0$ inherent in the 
underlying space. 
Thus, intuitively, UV-regularity and strict observational locality 
become more compatible than on ordinary geometry.
There exist `quasi-position representations' 
\cite{ak-jmp-qft,ak-gm-rm-prd,ak-hh-1},
built on maximal localisation states, which can be
used to establish the locality and causality properties 
of pointwise multiplications. A detailed
study on the special case $\Delta x_0>0,\Delta p_0=0$ (which allows 
a convenient momentum space representation) is in preparation
\cite{ak-gm-2}.
\sn
We remark that an alternative approach with a similar motivation, but
based on the canonical formulation of field theory is given in
\cite{doplicher}, see also \cite{grosse}.


\newpage
\begin{thebibliography}{**}
\bibitem{townsend} P. K. Townsend, Phys. Rev. D, Vol. 15, No 10:
2795-2801 (1976)
\bibitem{amati} D. Amati, M. Cialfaloni, G. Veneziano, Phys. Lett. 
B 216, 41 (1989)
\bibitem{maggiore} M. Maggiore, Phys. Lett. B 319: 83 (1993)
\bibitem{garay} L. J. Garay, Int. J. Mod. Phys. A10: 145 (1995)
\bibitem{ixtapa} A. Kempf, Proc. XXII DGM Conf. Sept.93 Ixtapa
(Mexico), Adv. Appl. Cliff. Alg (Proc. Suppl.) (S1):87 (1994)
\bibitem{ak-jmp-ucr} A. Kempf, J. Math. Phys. 35 (9): 4483 (1994)
\bibitem{ft} A. Kempf, Preprint DAMTP/94-33, hep-th/9405067                
\bibitem{prag} A. Kempf, Czech.J.Phys. (Proc. Suppl.) 44, 11-12:
1041 (1994)
\bibitem{ct} A. Kempf, Preprint DAMTP/96-06,
to appear in Proc. Intl. Workshop Lie Theory \& Applic. in Physics,
Clausthal, Aug 95, World Scientific (1996)
\bibitem{ak-jmp-qft} A. Kempf, DAMTP/96-22, hep-th/9602085
\bibitem{ak-gm-rm-prd} A. Kempf, G. Mangano, 
R.B. Mann, Phys. Rev. D52: 1108-1118 (1995) 
\bibitem{ak-hh-1} A. Kempf, H. Hinrichsen,
hep-th/9510144, in press in J. Math. Phys. (1996)
\bibitem{ak-gm-2} A. Kempf, G. Mangano, in preparation
\bibitem{connes} A. Connes, \it Noncommutative geometry, \rm AP (1994)
\bibitem{fdl} R.P. Feynman, 
\it Dirac Memorial Lecture, \rm CUP (1987)
\bibitem{doplicher} S. Doplicher, K. Fredenhagen, J. E. Roberts,
Phys. Lett. B331: 39 (1994)
\bibitem{grosse} H. Grosse, C. Klimcik, P. Presnadjer, hep-th/9510177, to
appear in Proc. Intl. Workshop on Lie Theory \& Applic. in Physics,
Clausthal, Aug 95, W.Sc. (1996)
\bibitem{gbook} J. Weidmann, \it Lin. Operatoren in Hilbertr{\"a}umen, \rm
(in German), Teubner (1976)
\bibitem{smirnov} V.I. Smirnov, \it A Course in Higher Math. V, \rm
Pergamon (1964)
\bibitem{ag} N.I. Akhiezer, I.M. Glazman, \it Theor. Lin. Oper. in 
Hilbert spaces, \rm Ungar (1963)
\bibitem{maurin} K. Maurin, \it Methods of Hilbert Spaces\rm, Polish 
Scientific Publishers (1967)
\bibitem{reedsimon} M. Reed, B. Simon, \it Fourier Analysis,
Self-Adjointness, \rm  AP (1975) 
\end{thebibliography}
\end{document}